\documentclass[reprint,aps,floatfix,footinbib,superscriptaddress]{revtex4-1}
\usepackage{color}
\usepackage[utf8]{inputenc}
\usepackage[T1]{fontenc}
\usepackage[usenames,dvipsnames]{pstricks}
\usepackage[colorlinks=true]{hyperref}
\usepackage{graphicx}
\usepackage{dcolumn}
\usepackage{bm}
\usepackage{amsmath,amssymb,amsthm} 
\usepackage{makeidx}
\usepackage{amsfonts}
\usepackage{comment}
\usepackage[usenames,dvipsnames]{pstricks}
\usepackage{subfigure}
\usepackage{braket}
\usepackage{float}

\begin{document}

\preprint{APS/123-QED}

\title{An edge-based formulation of elastic network models}

\author{Maxwell Hodges}\email{m.hodges14@imperial.ac.uk}
\author{Sophia N. Yaliraki}%
\affiliation{Department of Chemistry, Imperial College London, London~SW7~2AZ, United Kingdom}%
\author{Mauricio Barahona}\email{m.barahona@imperial.ac.uk}
\affiliation{Department of Mathematics, Imperial College London, London~SW7~2AZ, United Kingdom}%
\date{\today}

\begin{abstract}
We present an edge-based framework for the study of geometric elastic network models to model mechanical interactions in physical systems.  We use a formulation in the edge space, instead of the usual node-centric approach, to characterise edge fluctuations of geometric networks defined in $d$-dimensional space  and define the edge mechanical embeddedness, an edge mechanical susceptibility measuring the force felt on each edge given a force applied on the whole system. We further show that this formulation can be directly related to the infinitesimal rigidity of the network, which additionally permits three- and four-centre forces to be included in the network description. We exemplify the approach in protein systems, at both the residue and atomistic levels of description. 
\end{abstract}

\pacs{Valid PACS appear here}
\maketitle

\section{Introduction}
Elastic network models (ENMs) are ubiquitous in physics and have been applied to describe properties of a wide variety of structures including glasses~\cite{He1985, Palyulin2018}, biological tissue~\cite{Bischofs2008}, supercooled liquids~\cite{Yan2013} and, recently, the design of allosteric materials~\cite{Yan2016}.  A particularly useful application of ENMs, sparked by the seminal work of Tirion~\cite{Tirion1996}, has been in the study of protein structures, as the use of molecular dynamics (MD) simulations on biologically relevant timescales remains challenging.  The principal assumption of ENMs is that we may approximate the bottom of the potential energy well of a structure by a quadratic function, by taking the Taylor series of the potential energy with respect to node displacements about the minimum $\mathbf{r}_0$. In elastic models, the forces $\mathbf{f}$  are thus linear in the displacements $\mathbf{r}$, i.e., $\mathbf{f} = \mathbf{H(\mathbf{r}_0)} \, (\mathbf{r} - \mathbf{r}_0)$, where $\mathbf{H(\mathbf{r}_0)}$ is the Hessian matrix obtained by differentiating twice the potential function. Typically, the analysis of (infinitesimal) motions involves diagonalisation of $\mathbf{H}$ to determine the \emph{normal modes} of the protein.  Whilst real potential energy surfaces of proteins are complex, highly nonlinear and containing many minima~\cite{Henzler-Wildman2007a}, elastic models have been surprisingly effective for the analysis of slow equilibrium motions of proteins~\cite{Bastolla2014, Lopez-Blanco2014}.   Another common use of ENMs is for the calculation of node fluctuations, which have shown good agreement with crystallographic B-factors~\cite{Bahar1997, Yang2009}.  

The focus of ENMs has thus typically been on the node variables.  
Here, we present an \emph{edge} based formulation of ENMs, which instead puts the emphasis on the \emph{interactions} between the nodes, which in a mechanical framework corresponds to extensions (or compressions) of the `springs' associated with the edges. More formally, the edge changes are the \emph{dual} of the node motions~\cite{strang1986introduction}.  
An edge-centric approach has proved highly effective in previous studies of different networks~\cite{Amor2016, Schaub2014, Hodges2018}, and indeed the formulation presented need not be restricted to proteins and is general for networks embedded in $d$ dimensional space.  
There has been extended discussion in the literature over the use of networks with \emph{scalar} node variables to model 2- and 3-dimensional mechanical structures~\cite{Thorpe2007}.  By instead working in edge-space, we avoid this issue altogether since the scalar edge variables, which represent changes in edge length, appear naturally in the theory regardless of the dimensionality of the geometric structure.  
Historically, the Born-Huang model~\cite{born1954dynamical} has often formed the basis for the study of lattice structures but its weakness in handling disordered materials like glasses has been highlighted in the context of rigidity percolation~\cite{feng1984percolation} and more recently by Zaccone and Scossa-Romano~\cite{Zaccone2011}, who extended the Born model to include non-affine responses to external stresses.  
In many systems such as proteins, function is often driven by changes in structure, but crucially it is the \emph{relative} change in node positions that is of interest.  We thus show how to obtain  edge fluctuations in  elastic network models and compute the edge mechanical embeddedness as a useful property of the system. Finally, we show how this formulation naturally connects with the rigidity properties of the network, viewed as a set of edge constraints. We showcase the approach with specific protein examples.

\section{Theory}

\subsection{Edge-based Formulation of Geometric Elastic Network Models}
\begin{figure}
\centering
\includegraphics[width=0.35\textwidth]{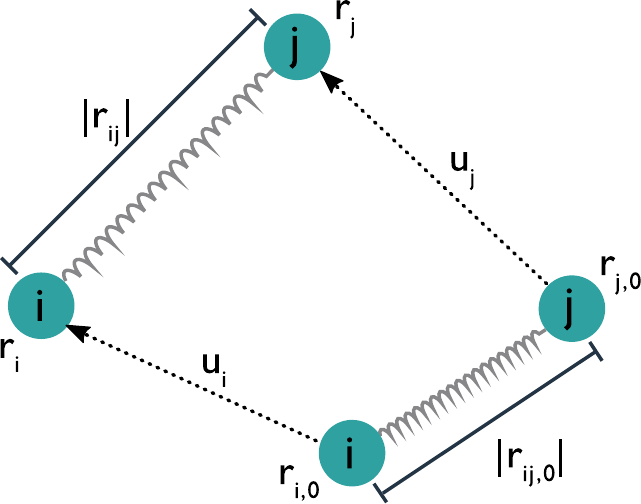}
\caption{The extension of the spring can be written in terms of the displacements of the nodes~\cite{strang1986introduction}} 
\label{ENM_displacements}
\end{figure}

Consider a network of $N$ nodes, associated with points in $d$-dimensional space $\mathbf{r}_{i,0} \in \mathbb{R}^d$, and with $E$ interactions between nodes (due to, e.g., physico-chemical potentials).
Let us denote the (small) node displacements as the $d$-dimensional vectors  
\begin{equation}
\mathbf{u}_i = \mathbf{r}_i - \mathbf{r}_{i,0}, \enskip i=1,\ldots, N.
\end{equation}
Each edge has an associated scalar variable $e_\alpha \in \mathbb{R}, \, \alpha=1, \ldots, E$, which measures its \emph{extension}, i.e., its change in length.  
The node and edge variables are related directly 
through the $N d \times E$ \textit{geometric incidence matrix}, $\mathcal{B}$: 
\begin{equation}
\mathbf{e} = \mathcal{B}^T \mathbf{u},
\label{eq:extension_displacement}
\end{equation}
where $\mathbf{u}^T=(\mathbf{u}^T_1, \ldots ,\mathbf{u}^T_N)$ is the $N d$-dimensional vector compiling the node displacements, and $\mathbf{e}$ is the $E$-dimensional vector of edge extensions.

To obtain the form of the geometric incidence matrix, note that each column of $\mathcal{B}$ is associated with an edge. Assuming small node displacements, it can be easily shown (Fig.~\ref{ENM_displacements} and Appendix~\ref{app:appendix_incidence}) that the extension of the edge $\alpha=(ij)$ between nodes $i$ and $j$ induced by the node displacements $\mathbf{u}$ (to order $\mathcal{O}(|\mathbf{u}|^2)$) is given by
\begin{align}
e_\alpha & = |\mathbf{r}_{ij}| - |\mathbf{r}_{ij,0}| \simeq \widehat{\mathbf{r}}^T_{ij} \, (\mathbf{u}_j - \mathbf{u}_i),
\label{eq:extension}
\end{align}
where $\widehat{\mathbf{r}}_{ij} = \mathbf{r}_{ij,0}/|\mathbf{r}_{ij,0}| =: \widehat{\mathbf{r}}_\alpha$ is the $d$-dimensional unit vector along the direction of edge $\alpha =(ij)$. 
Each row of $\mathcal{B}^T$ is a vector $\mathcal{B}_\alpha^T$ that follows from~\eqref{eq:extension_displacement}--\eqref{eq:extension}:
\begin{equation}
\begin{split}
\mathcal{B}_\alpha^T =
\begin{pmatrix}
0,&  \cdots, & 0, & -\widehat{\mathbf{r}}^T_\alpha, & 0, & \cdots, & 0, & \widehat{\mathbf{r}}^T_\alpha, & 0, & \cdots, 0
\end{pmatrix}, \\ 
\alpha = 1, \ldots E
\end{split}
\label{geometric_incidence}
\end{equation}
to form the geometric incidence matrix: 
$\mathcal{B} = 
    \begin{pmatrix}  
    \mathcal{B}_1 \cdots \mathcal{B}_E
    \end{pmatrix}$.
Note that the matrix $\mathcal{B}$ 
is akin to the standard $N \times E$ incidence matrix $B$ in graph theory~\cite{Schaub2014} but it includes full directional information through the $d$-dimensional edge unit vectors.

Invoking a mechanical description, we can use Hooke's Law and Newton's Third Law to obtain the usual linear relationship between input forces on the nodes $\mathbf{f}_{\text{nodes}}$ and the induced node displacements $\mathbf{u}$:
\begin{equation}
\mathbf{K}\mathbf{u} = \mathbf{f}_{\text{nodes}}
\end{equation}
where $\mathbf{f}_{\text{nodes}}$ is the $N d \times 1$ vector compiling the external forces on the nodes and $\mathbf{K}$ is the $N d \times N d$ stiffness matrix
\begin{equation}
\mathbf{K} = \mathcal{B} \mathbf{G} \mathcal{B}^T,
\label{eq:stiffness_bond}
\end{equation}
with $\mathbf{G}=\text{diag}(g_\alpha)$ denoting the $E \times E$ diagonal matrix of spring constants. The stiffness matrix is thus the Hessian of the system---indeed this is the only form the Hessian can take~\cite{Thorpe2007}.

Using our formulation, we can study the input-output properties of the system in terms of edge variables, i.e., the edge extensions $\mathbf{e}_{\text{out}}$ induced by \emph{external} forces $\bm{f}_{\text{in}}$ applied to the edges. 
Let us consider external forces applied along the edges, which we compile in an $E \times 1$ vector  $\bm{f}_{\text{in}}$. These edge forces result in edge compressions and stretches that induce forces on the nodes given by 
\[\mathbf{f}_{\text{nodes}} = \mathcal{B} \bm{f}_{\text{in}}.\]
We wish to disregard any components of the induced forces linked to \emph{rigid motions} of the elastic network, since such motions do not produce edge extensions ($\mathbf{e}_\text{out}=0$). This can be achieved naturally by considering the pseudo-inverse of the stiffness matrix. The induced non-rigid displacements are given by
\[ 
\mathbf{u}^{+} = \mathbf{K}^{+}\mathcal{B} \, \bm{f}_{\text{in}},
\]
where $\mathbf{K}^{+}$ is the Moore-Penrose pseudo-inverse of $\mathbf{K}$,
and the edge extensions induced by the applied edge forces are given by:
\begin{align}
\mathbf{e}_{\text{out}} = \mathcal{B}^T \mathbf{K}^{+}\mathcal{B} \, \bm{f}_{\text{in}} =: \mathbf{T} \, \bm{f}_{\text{in}}.
\label{edge_input_output}
\end{align}
For the input force $\bm{f}_{\text{in}}$, the output vector $\mathbf{e}_{\text{out}}$ records the induced change in length of all the edges in the network.
The meaning of the $E \times E$ matrix $\mathbf{T}$ is clear: given a unit force (input) applied along edge $\alpha$, the induced (output) extension at edge $\beta$ is the corresponding entry of $\mathbf{T}$:
\begin{align}
    e_\beta &= (\mathcal{B}_\beta)^T \mathbf{K}^{+} \mathcal{B}_\alpha = T_{\beta \alpha}.
    \label{effective_compliance}
\end{align}
As a consequence, the induced extension at the \emph{input} edge $i$ is given by the diagonal element $T_{\alpha \alpha}$, which, depending on the location of the spring within the network, might not necessarily be the same as if the spring was isolated.  This is the mechanical analogue of the \emph{effective resistance} in electrical networks~\cite{Ghosh2008,Schaub2014}, also known as the resistance distance~\cite{klein1993resistance},  
yet, in our case, it is both the connectivity and the \emph{geometry} of the network in $d$-dimensional space that determines edge responses. We exploit this concept in the following section through the definition of the edge mechanical embeddedness.  

\subsection{Edge Fluctuations and Mechanical Embeddedness}
One application of the model is to identify residue-residue interactions within a protein that exhibit the highest edge fluctuations.  To see this, consider the Langevin equation of a 3-dimensional elastic network ($d=3$) representing protein residues undergoing dynamical motion in a heat bath modelling the aqueous environment:
\begin{equation}
\mathbf{M}\frac{d^2 \mathbf{r}}{dt^2} + \mathbf{\Gamma}\frac{d\mathbf{r}}{dt} + \mathbf{K}(\mathbf{r} - \mathbf{r}_0) =  \bm{\eta} (t)
\end{equation}
where $\mathbf{M}$ is a diagonal mass matrix, $\mathbf{\Gamma}$ is the diagonal damping matrix and $\bm{\eta} (t)$ is a vector of i.i.d. Gaussian noises.  The damping terms arise from interactions of the protein with water and itself, and are typically large. Hence We consider the overdamped limit, where we may neglect inertial terms. Although larger damping is sometimes set for residues located deeper inside the structure~\cite{hinsen2000harmonicity}, for simplicity we set all damping values to be equal and we renormalise time to obtain: 
\begin{equation}
\frac{d\mathbf{r}}{dt} = - \mathbf{K}(\mathbf{r} - \mathbf{r}_0) + \bm{\eta} (t),
\label{brownian_dynamics}
\end{equation}
which has the general solution
\begin{equation}
\mathbf{u}(t) =
\mathbf{r}(t) - \mathbf{r}_0 = \int_{-\infty}^{t} \exp \left[ \mathbf{K}(t-s) \right] \bm{\eta} (s) \ ds, 
\end{equation}
where the residue position $\mathbf{r}(t)$ is now a random variable. We are again interested in the random fluctuations of the  \emph{edge extensions}~\eqref{eq:extension_displacement}. 
Utilising our geometric incidence matrix, one can show that the covariance matrix of the edge fluctuations is given by:
\begin{align}
\mathbb{E}\left[\mathbf{e}(t) \, \mathbf{e}(t)^T      
\right] 
&= \frac{1}{2} \mathcal{B}^T \mathbf{K}^{+} \mathcal{B}
= \frac{1}{2} \mathbf{T}
\label{edge_fluctuations}
\end{align}

In a number of papers, authors construct networks from residue-residue interactions and identify significant residues using measures of \emph{centrality}, such as edge or node betweenness~\cite{mcclendon2014dynamic, doncheva2012topological, Ribeiro2014}.  However, it is not clear what the physical significance of such measures is.  In contrast, the mean edge fluctuations are related to a graph theoretical measure called \emph{edge embeddedness}, first introduced in ~\cite{Schaub2014} in the context of random walks on networks and resistor networks.
We may then define the equivalent \emph{mechanical embeddedness} for edge $\alpha$ in a geometric elastic network in $d$-dimensions as:
\begin{equation}
    \varepsilon_\alpha = 1 - (\mathbf{G}\mathcal{B}^T \mathbf{K}^{+} \mathcal{B})_{\alpha \alpha} = 1 - g_\alpha T_{\alpha \alpha}
    \label{mechanical_embeddedness}
\end{equation}
The mechanical embeddedness has a clearer physical meaning: the second term is the fraction of the input force applied to edge $\alpha$ that edge $\alpha$ actually feels. If an edge feels all the force applied to it, it is not well "embedded' within the network and has a low value of $\varepsilon$  (i.e., it is not strongly coupled to the rest of the network and does not dissipate its fluctuations into the network). Conversely, edges that are more "embedded" within the network structure feel a lower force, dissipate fluctuations into the rest of the network, and have a larger $\varepsilon$ score nearer to 1. 

\subsection{Connection to Infinitesimal Rigidity}
There is also a straightforward relationship between the geometric incidence matrix $\mathcal{B}$ and the classic\emph{rigidity matrix} $\mathbf{R}$ of the structure, given in Eq.~\ref{eq:rigidity_definition}~\cite{Whiteley2005}.  In Appendix~\ref{app:rigidity}, we show that:
\begin{align}
\mathcal{B}^T = \mathbf{D}^{-1}\mathbf{R}, 
\end{align}
where $\mathbf{D}$ is the $E \times E$ diagonal matrix containing the interaction distances.  The rigidity matrix can be used to determine the rigid parts of the elastic network structure (i.e., those that allow no internal motion) and the flexible parts via the concept of \emph{infinitesimal rigidity}.  (The distinction between rigidity and infinitesimal rigidity is discussed in depth in Ref.~\cite{Whiteley2005}, but here we consider only \emph{generic structures} and so the two terms are equivalent.) The rigidity matrix of a three dimensional structure possesses six zero eigenvalues corresponding to three translations and three rotations, but may have additional zero eigenvalues associated with motions of the structure that lead to no change in the potential energy of springs in the network. 
In Appendix~\ref{app:Joao_algo}, we summarise an infinitesimal rigidity algorithm developed in~Ref.~\cite{JoaoThesis} that uses the set of eigenvectors associated with such additional zero eigenvalues (if they exist) to cluster the structure into rigid clusters.   At the cost of longer running times, this infinitesimal rigidity algorithm allows greater flexibility in the choice of constraints than the popular rigid cluster decomposition based on FIRST~\cite{Jacobs2001,jacobs2000computer}, which is computationally efficient, yet it imposes the presence of angle constraints in the network structure. .

In some systems, such as chemical bonds within molecules, we do in fact have additional constraints on the \emph{angles} between edges. Indeed, inclusion of three-centre interactions in the simulation of polymer glasses has been shown to be important for the interpretation of Raman scattering spectra~\cite{Milkus2018}.  We therefore consider three center interactions (and indeed four center interactions, corresponding to \emph{dihedral} angles).  Given three nodes $i, j, k$ with edges $(ij)$ and $(jk)$, we compute the change in length of edge $(ik)$ with the constraints that the other two edges are held constant: $|\mathbf{r}_{ij}|^2 = |\mathbf{r}_{ij,0}|^2$ and $|\mathbf{r}_{jk}|^2 = |\mathbf{r}_{jk,0}|^2$.
Expanding these equations and substituting into the expression for the extension of edge $(ik)$, which is opposite to node $j$, we obtain (see Appendix~\ref{app:three_centre}):
\begin{align}
  & e_{ik} = 
    \frac{1}{\left\vert \mathbf{r}_{ik,0} \right\vert}
    \begin{pmatrix}
    \mathbf{r}^T_{jk,0}, & (\mathbf{r}^T_{ij,0} - \mathbf{r}^T_{jk,0}), & -\mathbf{r}^T_{ij,0} 
    \end{pmatrix}
    \begin{pmatrix}
    \mathbf{u}_i \\ \mathbf{u}_j \\ \mathbf{u}_k 
    \end{pmatrix}    
\label{eq:three_centre}
\end{align}
Note that the three-centre extension~\eqref{eq:three_centre} relative to the `angle' at node $j$ is \emph{not} the same as if a two-center Hooke spring was placed between nodes $i$ and $k$.  From expressions of the form~\eqref{eq:three_centre}, we can construct the three-centre stiffness matrix $\mathbf{K}_{\text{angle}}$. 

Using a similar procedure, we also find the expression for the linear changes of a four-center interaction, by keeping the three two-center and two three-center interactions constant. Such changes lead to the four-centre stiffness matrix $\mathbf{K}_{\text{dihedral}}$. See Appendix~\ref{app:dihedral}. 

The total stiffness matrix is then the sum of the stiffness matrices: $\mathbf{K}_{\text{total}} = \mathbf{K}_{\text{bond}} + \mathbf{K}_{\text{angle}}+ \mathbf{K}_{\text{dihedral}}$, where $\mathbf{K}_{\text{bond}}$ is the two-centre matrix given in~\eqref{eq:stiffness_bond}.  The extensions $\mathbf{e}_{\text{out}}$ induced by input forces $\bm{f}_{\text{in}}$ follow the same form as in~\eqref{edge_input_output}:
\begin{equation}
\mathbf{e}_{\text{out}} = \mathcal{B}^T \mathbf{K}_{\text{total}}^{+}\mathcal{B} \, \bm{f}_{\text{in}} =:\mathbf{T}_{\text{total}}\, \bm{f}_{\text{in}}.
\label{eq:input_output_Ktotal}
\end{equation}
Below we study the effect of the different components of the stiffness matrix in the input-output properties of the system.

\section{Applications}

\subsection{A mechanical model of protein-ligand binding at the atomistic level}

Allostery is a biological process whereby the binding of a ligand to a protein leads to a functional change at a distant site (often the active site) of the protein~\cite{Tsai2014, Ribeiro2016}.  A common explanation for allostery is that ligand binding leads to a propagation of strain across the protein structure, potentially along specific residue pathways, causing a structural change at the active site.  

Here, we study this process using an atomistic elastic network model of a protein bound to an allosteric ligand. First, we measure the \emph{elastic response} elicited across the protein by the application of unit forces to all weak interactions between the ligand and the protein allosteric site with negative forces corresponding to \emph{compressions} of the source interactions and positive forces corresponding to \emph{extensions} (although the overall sign is arbitrary).
Furthermore, we apply infinitesimal rigidity analysis (Appendix~\ref{app:Joao_algo}) to obtain the rigid clusters within the protein to elucidate the propagation of the strain. Since strain cannot propagate through floppy regions, we expect both the allosteric site and active sites to be within the same rigid cluster if strain is to pass from one site to the other efficiently. 

Atomistic graphs are constructed from PDB files containing full 3D atomic data of protein structures, and the software FIRST~\cite{jacobs2000computer} to determine the presence of the various bond types (covalent, hydrogen and hydrophobic interactions).  We assign values to the spring constants of the edges with the correct order of magnitude, as per the Amber15fb force field~\cite{wang2017building} (Table~\ref{tab:bond_energies}). 
We do not use exact values for each interaction since it is difficult to assign spring constants to hydrogen bonds and hydrophobic interactions from force fields used in molecular dynamics. In such fields, hydrophobic interactions emerge from the presence of implicit or explicit water that favours interactions to polar regions of the protein, whereas hydrogen bonds are derived from electrostatic contributions.

\begin{table}[!htb]
\begin{center}
\caption {Springs constants for each of the elastic network interactions. \\}
\label{tab:bond_energies}
 \begin{tabular}{c  c } 
 \hline
 Interaction & Spring constant\\ [0.5ex] 
 \hline
 Covalent & 100\\ 
 Hydrogen &  10\\
 Hydrophobic & 1\\
 Angle & 1\\
 Dihedral & 0.1\\
\hline
\end{tabular}
\end{center}
\end{table}

\begin{figure}[htb!]
\centering
\includegraphics[width=0.5\textwidth]{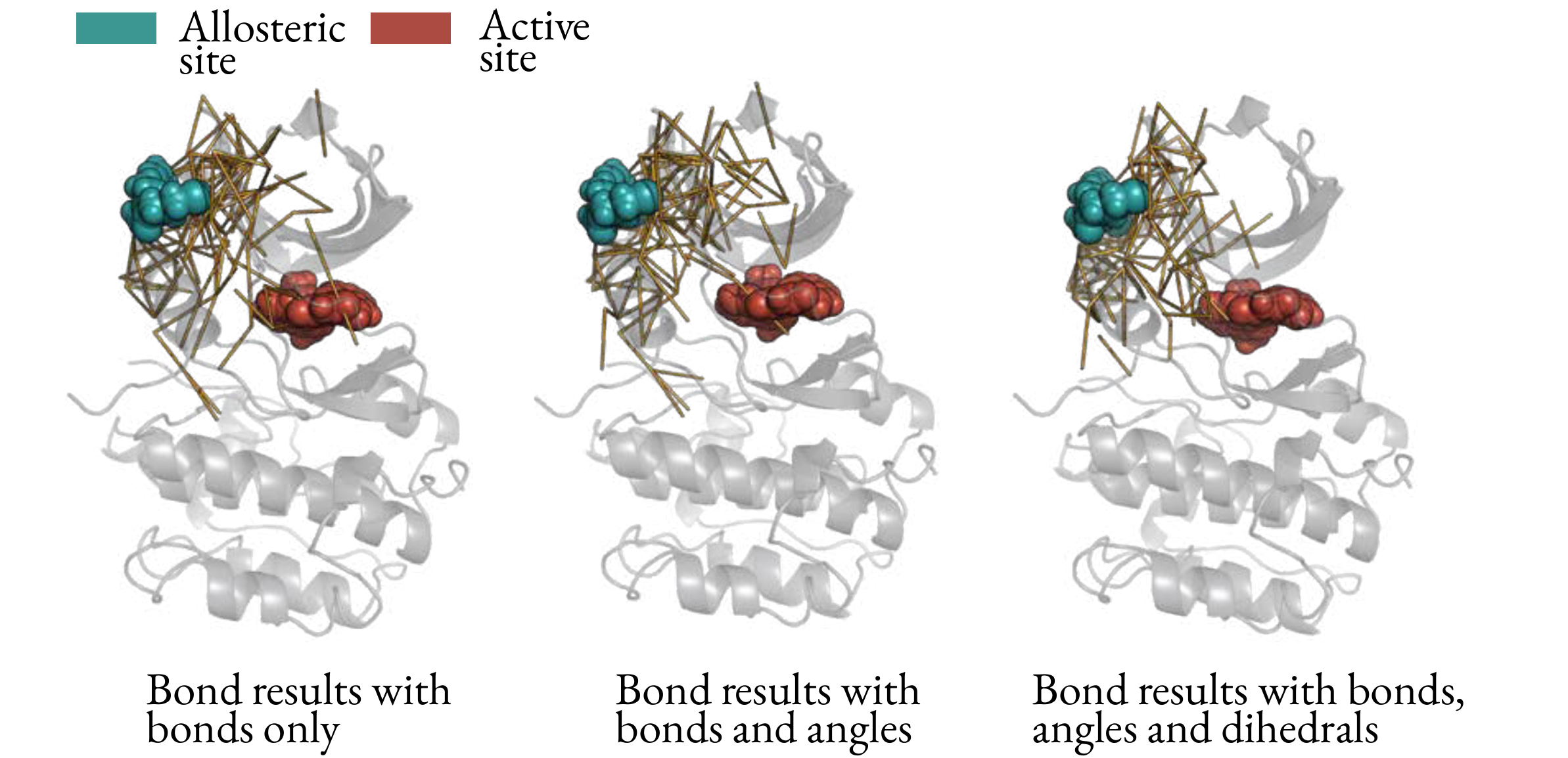}
\caption[Elastic response of PDK1]{Elastic response of PDK1 (PDB code: 3ORZ~\cite{sadowsky2011turning}). The top 2\% of bonds by absolute extension are shown for three cases: (left) only the 2-centre interactions (bonds) are included in the network; (centre) 3-centre (angle) interactions are added to the network and the top \emph{output extensions} of bonds are shown; (right) dihedral angles are added and again only top output bond displacements are shown.}
\label{PDK1_elasticresponse}
\end{figure}

\begin{figure}[htb!]
\centering
\includegraphics[width=0.49\textwidth]{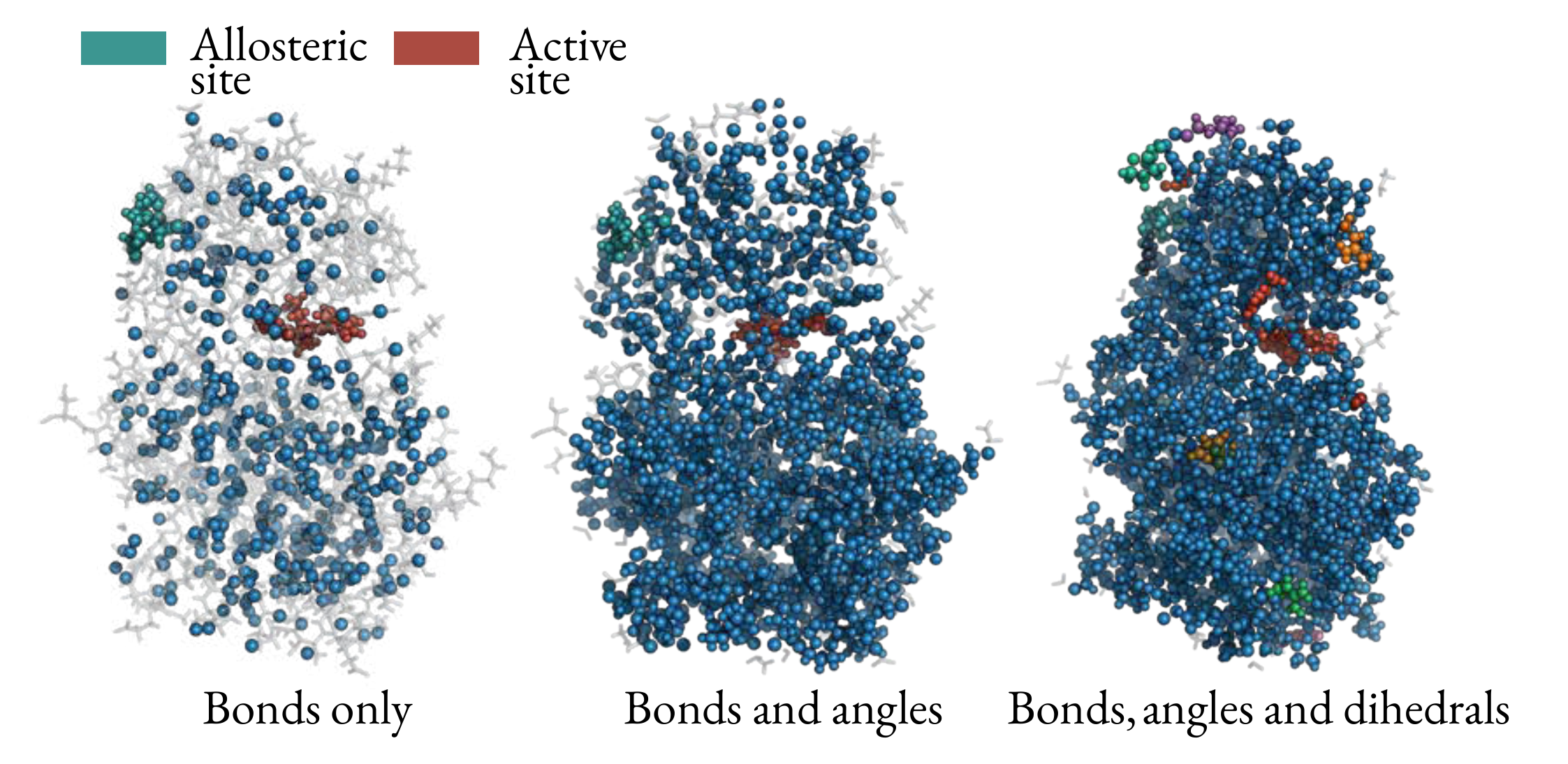}
\caption{Infinitesimal rigidity of PDK1. Each cluster has a different colour and `floppy' (non-rigid) atoms are shown in transparent grey:  (left) only bonds included as constraints, leading to a single large cluster in blue with all other atoms floppy; (centre) angle constraints added;  (right) dihedral constraints also added leading to a large rigid cluster extending over the whole protein. The rigid clusters and floppy atoms are computed using the algorithm in Appendix~\ref{app:Joao_algo}, which was introduced in~\cite{JoaoThesis}.}
\label{PDK1_infrig}
\end{figure}

We obtain the output extension for all edges in the protein in response to inputs at source edges given by the interactions with the ligand, and exemplify the results through the allosteric protein PDK1 (PDB code: 3ORZ). 
In Fig.~\ref{PDK1_elasticresponse} we show the top $2\%$ of bonds by \emph{absolute} length change (i.e., we do not discriminate between bond stretching or compression) in three scenarios: (i) where only the two-centre bond interactions are used to construct the elastic network, as is traditionally the case with elastic networks of proteins;  (ii) where angle constraints between pairs of covalent bonds are included; and (ii) where dihedral angle constraints from double bonds are also modelled.  Given that the highest scoring interaction in the bonds-only network (i)  (the hydrophobic interaction between Lys120 and Asn122) exhibits an extension of 0.766, we choose to represent the top $2\%$, which exhibit changes above 0.01, as a reasonable cutoff.
The most stretched edges are all located in the area connecting the allosteric and active sites. Furthermore, the infinitesimal rigidity analysis (Fig.~\ref{PDK1_infrig}) shows that, even for the 2-centre stifness matrix, the allosteric site and the region around the active site (Val96, Lys111, Tyr161, Ala162, Thr222, Asp223) all appear in a rigid cluster, with Leu88 the only active site residue that has no atoms within the rigid cluster.  When 3-centre and 4-centre constraints are included, the protein becomes strongly, due to the qualitative nature of the infinitesimal rigidity condition. It appears then plausible that propagation of strain may be emitted from binding at the allosteric site towards the active site, particularly through the rigid cluster formed by the 2-centre interactions that contains a smaller subset of the atoms in the protein.

The results of the elastic response show that the $\mathbf{e}_\text{out}$ decrease exponentially with distance (correlation coefficient = -0.603), 
even when angles and dihedrals are included (see Fig.~\ref{fig:exponential_decay} in the Appendix). Such a response is similar to random networks~\cite{Yan2016} and is not suggestive of a structure exclusively optimised for directed perturbations.
Indeed, the two largest extensions are found in the Lys120-Asn122 (0.766) and Val124-Pro125 (0.437) hydrophobic interactions, which are within $5~\mbox{\AA}$ of the allosteric source site, whereas the active site is around $17~\mbox{\AA}$ away.  
The highest scoring interactions involving active site residues are two Lys111-Phe157 hydrophobic interactions, which have extensions of 0.0122, and rank $130^{\text{th}}$ and $131^{\text{st}}$.  Of the top $2\%$ interactions (149 out of a total of 7391 edges) by output extensions, all but 4 are hydrophobic interactions. This is unsurprising given they have the weakest spring constants, but appears to lead to those weak interactions near the allosteric site effectively acting like a sponge, absorbing the shock of input forces and preventing long-range transfer of displacement.  If we change the force constant of the hydrophobic interactions to 10 (the same as the hydrogen bonds), the range of the propagation increases.  However, it is difficult to rationally assign such spring constant values to the hydrophobic interactions, and  there does not appear to be strong evidence that the allosteric effect exhibited by PDK1 is mediated by traversal of strain energy.  We have performed the elastic response analysis on a further two proteins (h-Ras, ATCase) with similar results. Hence our examples indicate that topological notions alone (such as rigidity) do not fully determine if a mechanical explanation for allostery is plausible, as the particular values of the edge spring constants are also crucial.

\subsection{Fluctuations of residue-residue interactions}

We applied our edge-based geometric formulation to a residue-residue interaction network (RRIN), i.e., an elastic network model of a protein at the residue level. We constructed several RRINs for ADK (4AKE~\cite{Muller1996})  using different distance cutoffs ($7 \mbox{\AA}, 10 \mbox{\AA}, 12 \mbox{\AA}, 15 \mbox{\AA}$) and obtained the average displacement for each of the edge interactions~\eqref{edge_fluctuations}. To decide on the appropriate cutoff, we computed Spearman's correlation coefficient ($\rho$) of the resulting extensions across the RRINs created with different cut-offs and found greater robustness for larger cutoff values: $\rho = 0.216$ between the $7 \mbox{\AA}$ and $10 \mbox{\AA}$ RRINs; increasing to $\rho = 0.679$, between the $10 \mbox{\AA}$ and $12 \mbox{\AA}$ RRINs; and increasing further to $\rho = 0.801$ between the RRINs created with $12 \mbox{\AA}$ and $15 \mbox{\AA}$ cutoffs.  (Below $7 \mbox{\AA}$, zero energy modes appear in the network as revealed by singular value decomposition of the rigidity matrix $R$.)  We thus use a cutoff of $12 \mbox{\AA}$ here and a single force constant (arbitrarily set to 1), in line with other reports in the literature~\cite{Atilgan2001}. 

\begin{figure}[htb!]
\centering
\includegraphics[width=0.49\textwidth]{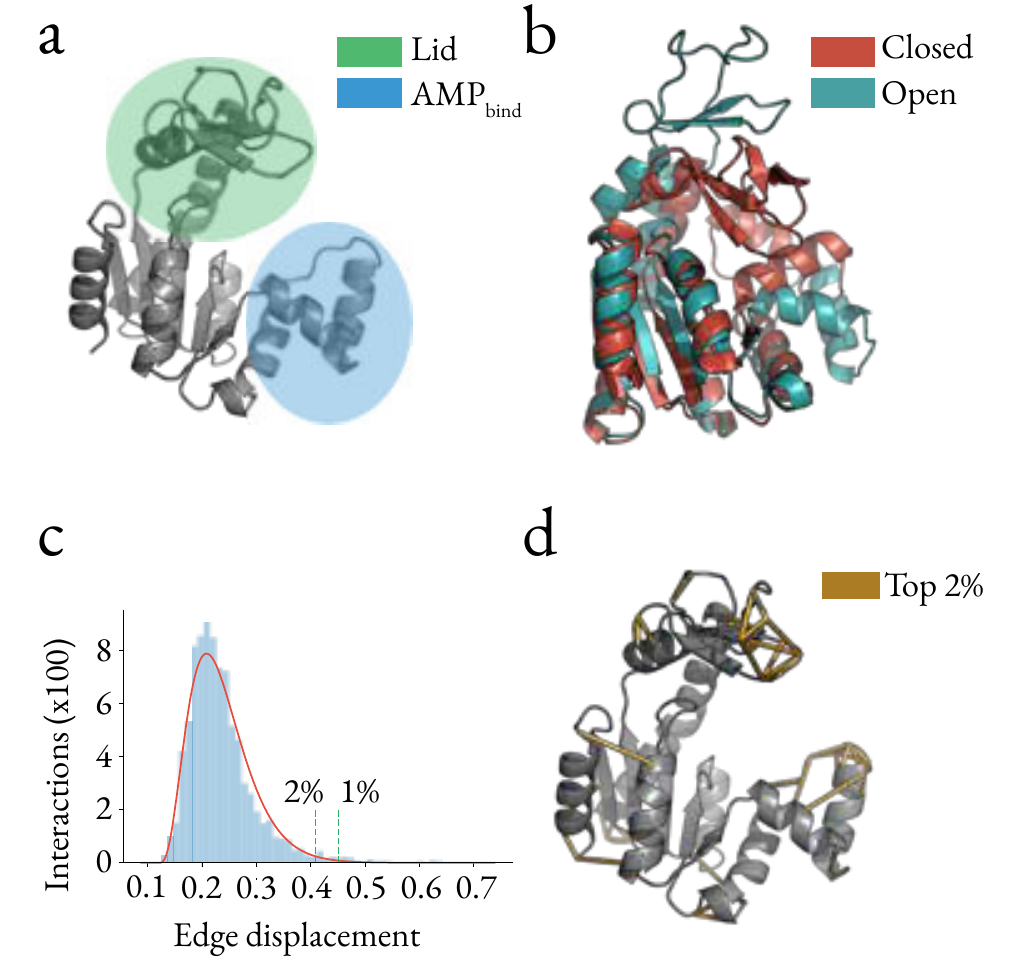}
\caption{(a) Structure of ADK from \textit{Escherichia coli} (open conformation, PDB: 4AKE), with the lid and AMP$_\text{bind}$ domains highlighted. (b) Closed (1AKE) and open (4AKE) forms of ADK showing that the main differences are in the lid and AMP$_\text{bind}$ domains. (c) The distribution of average edge displacements computed for open ADK (4AKE).  (d) The top 2\% residue-residue interactions with highest displacements are concentrated in the lid and AMP$_\text{bind}$ domains. Note that the viewpoint of the structures in (b) has been changed slightly relative to (a) and (d), so as to facilitate the visualisation of the differences between the open and closed conformations.} 
\label{ADK_results}
\end{figure}

Fig.~\ref{ADK_results} shows the results of our analysis of the this RRIN of open ADK (4AKE). A relatively right-skewed distribution of edge displacement values is observed (Pearson median skewness = 0.580), with average value of $0.236$ and a number of interactions scoring significantly highly.  The top $2\%$ of interactions by rank are those scoring above $0.409$ and the top $1\%$ score above $0.452$.  The most highly scoring interactions are clustered primarily in the lid and AMP$_{\text{bind}}$ domains, corresponding closely to those regions of the protein that are structurally altered during the open-to-closed transition.  Qualitatively similar observations were obtained by Mitchell~\textit{et al}~\cite{Mitchell2016} from dynamic data, i.e., by comparing residue displacements across an NMR ensemble of structures to calculate local strain.  Note that here, however, just a \emph{single} structure is used and strain is predicted \textit{a priori}, emphasising the fact that the intrinsic topology of the protein determines where strain is distributed to assist function.  The highest scoring interaction with 0.701 is Gly56-Lys57; Lys57 is one of the residues that shifts more than $10\mbox{\AA}$ during the open-to-closed transition~\cite{Gunther2003} whilst Gly56 has been shown to display particularly high fluctuations in coarse-grained MD simulations at the residue level~\cite{Wang2018}. Since we follow standard convention and use the same force constant for all residue-residue interactions (arbitrarily set to 1), by Eq.~\eqref{mechanical_embeddedness} (with $G$ equal to the identity matrix) we can see that the interactions with the greatest average extension are also those with the lowest \emph{mechanical embeddedness}, demonstrating a conceptual link to the network theory interpretation of protein structure.

\section{Conclusions}

In conclusion, we have presented here a framework for the study of geometric elastic network models in $d$-dimensional space through an alternative formulation in the edge space. The edge space is often more natural than the dual node space, as it allows the direct description of interactions and constraints, with their associated energies (or costs).  By conveniently working with the internal coordinates of the network, there is no need to consider rigid motions or to arbitrarily "pin" nodes. In many systems, such as proteins, it is changes in the interactions that are of interest rather than the nodes themselves, and optimization problems involving edge variables are more naturally dealt with using an edge-based framework. 
 
\section{Acknowledgements}
 This work was funded by an EPSRC Centre for Doctoral Training PhD Studentship from the Institute of Chemical Biology (Imperial College London) awarded to MH. MB acknowledges funding from the EPSRC project EP/N014529/1 supporting the EPSRC Centre for Mathematics of Precision Healthcare. We thank Joao Costa for his help and insight with the infinitesimal rigidity algorithm developed in his PhD thesis.

\bibliography{embeddedness}

\newpage
\pagebreak

\appendix

\section{Derivation of the geometric incidence matrix for the two-centre interactions}
\label{app:appendix_incidence}

For a spring connecting two nodes $i$ and $j$ with initial positions $\mathbf{r}_{i,0}$ and $\mathbf{r}_{j,0}$ and final positions $\mathbf{r}_{i}$ and $\mathbf{r}_{j}$, let $\mathbf{u}_i$ and $\mathbf{u}_j$ be the displacement of each node.  We denote the final node positions by:
\begin{align}
    \mathbf{r}_{i} &= \mathbf{r}_{i,0} + \mathbf{u}_{i} \nonumber \\
    \mathbf{r}_{j} &= \mathbf{r}_{j,0} + \mathbf{u}_{j}
\end{align}
with the vectors describing the springs before and after the extension (or compression) being:
\begin{align}
    \mathbf{r}_{ij,0} &= \mathbf{r}_{j,0} - \mathbf{r}_{i,0} \nonumber \\
    \mathbf{r}_{ij} &= \mathbf{r}_{j} - \mathbf{r}_{i}
\end{align}

Consider the extension of an edge written in terms of the displacements of its associated nodes:  
\begin{equation}
e_{ij} = |\mathbf{r}_{ij}| - |\mathbf{r}_{ij,0}|.
\label{edge_displacement}
\end{equation}
We first expand $|\mathbf{r}_{ij,0}|$:
\begin{align}
    \left\vert \mathbf{r}_{ij,0} \right\vert^2 &= \mathbf{r}_{ij,0} \cdot \mathbf{r}_{ij,0} \nonumber \\
    &= \left( \mathbf{r}_{j,0} - \mathbf{r}_{j,0} \right) \cdot \left( \mathbf{r}_{j,0} - \mathbf{r}_{j,0} \right)
    \label{initial_position}
\end{align}
then expand $|\mathbf{r}_{ij}|$:
\begin{align}
    \left\vert \mathbf{r}_{ij} \right\vert^2 &= \mathbf{r}_{ij} \cdot \mathbf{r}_{ij}  
     = \left \vert \mathbf{r}_{j,0} + \mathbf{u}_{j} - \mathbf{r}_{i,0} - \mathbf{u}_{i}  \right \vert ^2   \nonumber \\    
&= \left \vert (\mathbf{r}_{j,0} - \mathbf{r}_{i,0}) + (\mathbf{u}_j - \mathbf{u}_i) \right \vert^2   \nonumber
\end{align}

Using Eq.\eqref{initial_position}, we substitute terms:
\begin{align}
    \left\vert \mathbf{r}_{ij} \right\vert^2 &= \left\vert \mathbf{r}_{ij, 0} \right\vert^2 + 2 \ \mathbf{r}_{ij, 0} \left(\mathbf{u}_j - \mathbf{u}_i \right) \nonumber \\
    &= \left\vert \mathbf{r}_{ij, 0} \right\vert^2 + 2 \ \left\vert \mathbf{r}_{ij,0} \right\vert \ \frac{\mathbf{r}_{ij, 0}}{\left\vert \mathbf{r}_{ij,0} \right\vert} \left(\mathbf{u}_j - \mathbf{u}_i \right)
\end{align}
We can complete the square:
\begin{equation}
    \left\vert \mathbf{r}_{ij} \right\vert^2 = \left(\left\vert \mathbf{r}_{ij, 0} \right\vert + \frac{\mathbf{r}_{ij, 0}}{\left\vert \mathbf{r}_{ij,0} \right\vert} \left(\mathbf{u}_j - \mathbf{u}_i \right)  \right)^2 + \Theta(\mathbf{u}^2)
\end{equation}
and make a linear approximation by dropping nonlinear terms and square rooting both sides:
\begin{align}
    \left\vert \mathbf{r}_{ij} \right\vert = \left\vert \mathbf{r}_{ij,0} \right\vert + \frac{\mathbf{r}_{ij, 0}}{\left\vert \mathbf{r}_{ij,0} \right\vert} \left(\mathbf{u}_j - \mathbf{u}_i \right).
\end{align}
By referring to Eq.\eqref{edge_displacement} we now have an expression for the change in spring length in terms of the node displacements:
\begin{equation}
    e_{ij} = \frac{1}{\left\vert \mathbf{r}_{ij,0} \right\vert}  \mathbf{r}_{ij, 0} \cdot \left(\mathbf{u}_i - \mathbf{u}_j \right),
\end{equation}
which can be expressed in vector form as:
\begin{equation}
     e_{ij} = \frac{1}{\left\vert \mathbf{r}_{ij,0} \right\vert}
     \left[ \begin{pmatrix}    1, & -1  \end{pmatrix} \otimes \mathbf{r}^T_{ij, 0} \right]
    \begin{pmatrix}
    \mathbf{u}_i \\ \mathbf{u}_j
    \end{pmatrix} 
\end{equation}
Here we have shown the expression for an isolated spring, For a spring in a network, the elements of the vectors relating to nodes not involved with the spring would be zero so that our geometric incidence matrix $\mathcal{B}$ has rows of the form~ \eqref{geometric_incidence}.

\section{Derivation of the three centre terms}
\label{app:three_centre}

\begin{figure}[!htb]
\centering
\includegraphics[width=0.25\textwidth]{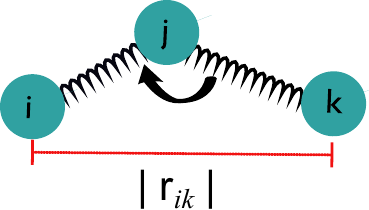}
\caption[Angle interaction]{Schematic for the derivation of the three-centre (angle) interaction, where the two-centre bond lengths $(ij), (jk)$ are kept constant and we compute the extension $i-k$ under those constraints.}
\label{three_centre}
\end{figure}

We wish to find an expression for the change in length of the distance $i - k$ 
\begin{equation}
    e_{ik} = \left\vert \mathbf{r}_{ik} \right\vert - \left\vert \mathbf{r}_{ik,0} \right\vert
    \label{extension}
\end{equation}
in terms of the node displacements $\mathbf{u}_i, \mathbf{u}_j \text{ and } \mathbf{u}_k$ under the assumption that the edge distances are fixed (Fig.~\ref{three_centre}).
The initial distance, using the cosine rule, is:
\begin{equation}
    \left\vert \mathbf{r}_{ik,0} \right\vert^2 = \left\vert \mathbf{r}_{ij,0} \right\vert^2 + \left\vert \mathbf{r}_{jk,0} \right\vert^2 - 2 \left\vert \mathbf{r}_{ik,0} \right\vert \left\vert \mathbf{r}_{jk,0} \right\vert \cos \theta_0
\label{inital_distance}
\end{equation}
and likewise the distance after the perturbation of the three nodes is:
\begin{equation}
    \left\vert \mathbf{r}_{ik} \right\vert^2 = \left\vert \mathbf{r}_{ij} \right\vert^2 + \left\vert \mathbf{r}_{jk} \right\vert^2 - 2 \left\vert \mathbf{r}_{ik} \right\vert \left\vert \mathbf{r}_{jk} \right\vert \cos \theta
\label{final_distance}
\end{equation}

We apply the constraints $\left\vert \mathbf{r}_{ij} \right\vert = \left\vert \mathbf{r}_{ij,0} \right\vert$ and $\left\vert \mathbf{r}_{jk,0} \right\vert = \left\vert \mathbf{r}_{jk,0} \right\vert$ as we are interested only in the change in angle, not in any two centre changes.  We can then substitute terms from~\eqref{inital_distance} into~\eqref{final_distance}:
\begin{align}
    \left\vert \mathbf{r}_{ik} \right\vert^2 &= \left\vert \mathbf{r}_{ik,0} \right\vert^2 + 2 \left\vert \mathbf{r}_{ik,0} \right\vert \left\vert \mathbf{r}_{jk,0} \right\vert \cos \theta_0 - 2 \left\vert \mathbf{r}_{ik,0} \right\vert \left\vert \mathbf{r}_{jk,0} \right\vert \cos \theta \nonumber \\
    &= \left\vert \mathbf{r}_{ik,0} \right\vert^2 - 2 \left\vert \mathbf{r}_{ik,0} \right\vert \left\vert \mathbf{r}_{jk,0} \right\vert \left( \cos \theta - \cos \theta_0 \right)
\end{align}
We can rewrite this expression as:
\begin{equation*}
    \left\vert \mathbf{r}_{ik} \right\vert^2 = \left\vert \mathbf{r}_{ik,0} \right\vert^2 - 2\left\vert \mathbf{r}_{ik,0} \right\vert \left( \frac{\left\vert \mathbf{r}_{ij,0} \right\vert \left\vert \mathbf{r}_{jk,0} \right\vert}{\left\vert \mathbf{r}_{ik,0} \right\vert} \left( \cos \theta - \cos \theta_0 \right) \right)
\end{equation*}
so that we can complete the square:
\begin{equation*}
    \left\vert \mathbf{r}_{ik} \right\vert^2 = \left( \left\vert \mathbf{r}_{ik} \right\vert - \frac{\left\vert \mathbf{r}_{ij,0} \right\vert \left\vert \mathbf{r}_{jk,0} \right\vert}{\left\vert \mathbf{r}_{ik,0} \right\vert} \left( \cos \theta - \cos \theta_0 \right) \right)^2 + \Theta\left(\left\vert \mathbf{r} \right\vert\right)^2
\end{equation*}
By ignoring nonlinear terms and square rooting both sides can then write our extension from Eq.\eqref{extension} in terms of the initial and final angles:
\begin{align*}
     e_{ik} = \left(\frac{\left\vert \mathbf{r}_{ij,0} \right\vert \left\vert \mathbf{r}_{jk,0} \right\vert}{\left\vert \mathbf{r}_{ik,0} \right\vert} \left( \cos \theta - \cos \theta_0 \right) \right)
\end{align*}

However, we wish to derive the extension in terms of node displacements (in Cartesian coordinates) and so substitute using the definition of the dot product:
\begin{equation}
    e_{ik} = \frac{1}{\left\vert \mathbf{r}_{ik} \right\vert} \left( \left( \mathbf{r}_{ij} \cdot \mathbf{r}_{jk} \right) -  \left( \mathbf{r}_{ij,0} \cdot \mathbf{r}_{jk,0} \right) \right)
    \label{difference_dot_products}
\end{equation}
where we have again used the fact that the two bonds have not changed length.  We now define the displacements of the nodes in terms of the bond vectors before and after perturbation:
\begin{align}
    &\mathbf{u}_i = \mathbf{r}_i - \mathbf{r}_{i,0} \nonumber\\
    &\mathbf{u}_j = \mathbf{r}_j - \mathbf{r}_{j,0} \nonumber\\
    &\mathbf{u}_k = \mathbf{r}_k - \mathbf{r}_{k,0}
    \label{displacements}
\end{align}

We now expand out Eq.\eqref{difference_dot_products}:
\begin{align}
    e_{ik} &= \frac{1}{\left\vert \mathbf{r}_{ik} \right\vert} \left(   \mathbf{r}_{i} \cdot \mathbf{r}_{j} - \mathbf{r}_{i} \cdot \mathbf{r}_{k} - \mathbf{r}_{j} \cdot \mathbf{r}_{j} + \mathbf{r}_{j} \cdot \mathbf{r}_{k} \right. \nonumber \\
    &- \left. \mathbf{r}_{i,0} \cdot \mathbf{r}_{j,0} + \mathbf{r}_{i,0} \cdot \mathbf{r}_{k,0} + \mathbf{r}_{j,0} \cdot \mathbf{r}_{j,0} - \mathbf{r}_{j,0} \cdot \mathbf{r}_{k,0}
    \right) \label{dot_products_expanded}
\end{align}

Substituting terms from~\eqref{displacements}, we drop the nonlinear terms in the second line of~\eqref{dot_products_expanded} to get:
\begin{align*}
   e_{ik} =  \ &(\mathbf{r}_{i,0} + \mathbf{u}_i) \cdot (\mathbf{r}_{j,0} + \mathbf{u}_j) 
        -(\mathbf{r}_{i,0} + \mathbf{u}_i) \cdot (\mathbf{r}_{k,0} + \mathbf{u}_k) \nonumber \\
        -&(\mathbf{r}_{j,0} + \mathbf{u}_j) \cdot (\mathbf{r}_{j,0} + \mathbf{u}_j) 
        +(\mathbf{r}_{j,0} + \mathbf{u}_j) \cdot (\mathbf{r}_{k,0} + \mathbf{u}_k) 
\end{align*}
Now we again drop nonlinear terms that result from the expansion of each of the dot products to give:
\begin{align*}
     e_{ik} = \ \frac{1}{\left\vert \mathbf{r}_{ik} 
     \right\vert} \ & \left ( \mathbf{r}_{i,0} \cdot \mathbf{u}_{j} + \mathbf{r}_{j,0} \cdot \mathbf{u}_{i} 
    - \mathbf{r}_{i,0} \cdot \mathbf{u}_{k} - \mathbf{r}_{k,0} \cdot \mathbf{u}_{i} \right. \nonumber  \\
    -   & \left. \mathbf{r}_{j,0} \cdot \mathbf{u}_{j} - \mathbf{r}_{j,0} \cdot \mathbf{u}_{j} 
    + \mathbf{r}_{j,0} \cdot \mathbf{u}_{k} + \mathbf{r}_{k,0} \cdot \mathbf{u}_{j} \right)
\end{align*}
which can be written more compactly as:
\begin{align}
    e_{ik} &= \left( \mathbf{r}_{j,0} - \mathbf{r}_{k,0} \right) \mathbf{u}_{i} \nonumber \\
    &- \big( \left( \mathbf{r}_{j,0} - \mathbf{r}_{i,0} \right) + \left( \mathbf{r}_{j,0} - \mathbf{r}_{k,0} \right) \big) \mathbf{u}_{j} \nonumber \\
    &+ \left( \mathbf{r}_{j,0} - \mathbf{r}_{i,0} \right) \mathbf{u}_{k},
\end{align}
or, in vector form, as:
\begin{equation}
    \begin{split}
    \label{eq:three_centre_expression} 
    & e_{ik} = 
    \frac{1}{\left\vert \mathbf{r}_{ik,0} \right\vert}
    \begin{pmatrix}
    \mathbf{r}^T_{jk,0}, & (\mathbf{r}^T_{ij,0} - \mathbf{r}^T_{jk,0}),  & -\mathbf{r}^T_{ij,0} 
    \end{pmatrix}
    \begin{pmatrix}
    \mathbf{u}_i \\ \mathbf{u}_j \\ \mathbf{u}_k 
    \end{pmatrix} 
  \end{split}
\end{equation}

 As for the two-centre case, each row of $\mathcal{B}_{\text{angle}}$, the geometric incidence  matrix for the three-center interactions, has the form~\eqref{eq:three_centre_expression}  but with zeros in the entries relating to nodes not involved in the corresponding interaction. 
    
The stiffness matrix for the three centre interactions can then be constructed similarly: 
\[\mathbf{K}_{\text{angle}} = \mathcal{B}_{\text{angle}} \mathbf{G}_{\text{angle}} \mathcal{B}_{\text{angle}}^T, \] 
with $\mathbf{G}_{\text{angle}}$ the diagonal matrix containing the spring constants for the three-centre (`angle') interactions.

\section{Expression for the four centre terms}
\label{app:dihedral}
\begin{figure}[!htb]
\centering
\includegraphics[width=0.4\textwidth]{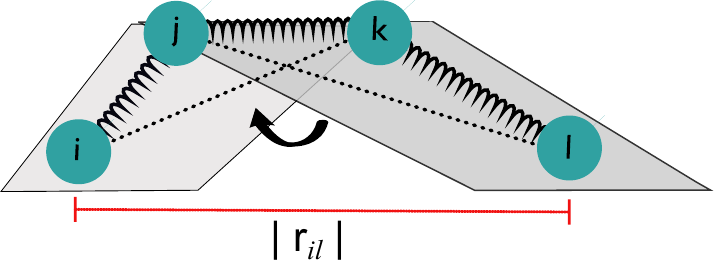}
\caption[Dihedral interaction]{Schematic for the derivation of the dihedral interaction: the three two centre bond lengths $(ij), (jk), (kl)$ remain constant, as well as the two three-centre angle interactions $(ijk)$ and $(jkl)$ marked with the dashed lines. We compute the extension $i-l$ under those constraints.}
\label{dihedral}
\end{figure}

The derivation for the four centre (or dihedral) terms is similar to the three centre case, where we compute the extension in the length between $i-l$ while keeping constant the two angular terms $(ijk)$ and $(jkl)$, as well as each of the three bond lengths $(ij), (jk), (kl)$ (Fig.~\ref{dihedral}).
This leads to the expression:
\begin{equation}
    \begin{split}
    e_{il} = & \frac{1}{ \left\vert \mathbf{r}_{il,0} \right\vert} 
    \left[
    \left(\mathbf{r}_{jl,0} + \mathbf{r}_{ik,0} \right) \cdot \mathbf{u}_{i}
    +\left(\mathbf{r}_{ij,0} + \mathbf{r}_{jk,0} + \mathbf{r}_{jl,0} \right) \cdot \mathbf{u}_{j}  \right. \\
     & -\left(\mathbf{r}_{jk,0} + \mathbf{r}_{kl,0} + \mathbf{r}_{ik,0} \right) \cdot \mathbf{u}_{k}  
    -\left(\mathbf{r}_{jl,0} + \mathbf{r}_{ik,0} \right) \cdot \mathbf{u}_{l} ], 
    \label{eq:Max_fourcentre}
\end{split}
\end{equation}
which, by using $\mathbf{r}_{jl,0} = \mathbf{r}_{jk,0} + \mathbf{r}_{kl,0}$ and $\mathbf{r}_{ik,0} = \mathbf{r}_{ij,0} + \mathbf{r}_{jk,0}$, is transformed into:
\begin{align}
    e_{il} = 
    \frac{1}{\left\vert \mathbf{r}_{il,0} \right\vert}
    \left[
     \begin{pmatrix}
    1, & 1, &  -1,  & -1 
    \end{pmatrix}
    \otimes
    \left(\mathbf{r}^T_{ij,0}+2 \mathbf{r}^T_{jk,0} + \mathbf{r}^T_{kl,0} \right)
    \right]
    \begin{pmatrix}
    \mathbf{u}_i \\ \mathbf{u}_j \\ \mathbf{u}_k  \\  \mathbf{u}_l 
    \end{pmatrix} .
    \label{eq:MB_fourcentre}
\end{align}

\section{Definition of the rigidity matrix}
\label{app:rigidity}
Here we summarise standard calculations covered in Refs.~\cite{Whiteley2005, JoaoThesis, costa2006role}.

In this Appendix~\ref{app:rigidity} and the subsequent Appendix~\ref{app:Joao_algo}, we adopt the usual notation in rigidity theory, where the positions of the points are represented as $\mathbf{p}_i$, and the system is defined by a set $\mathcal{M}$ of $M$ distance constraints. Note that in the rest of the paper, $\mathbf{p}_i$  are denoted as $\mathbf{r}_i$, and the $M$ constraints correspond to the $E$ edges of the graph encapsulating the interactions.

The rigidity problem for $N$ points $\mathbf{p}_i \in \mathbb{R}^d, \, i=1,\dots,N$ with a set $\mathcal{M}$ of distance constraints $c_\alpha, \, \alpha=1, \dots, M$ is given explicitly by the following set of $M$ equations:
\begin{equation}
|\mathbf{p}_i - \mathbf{p}_j|^2 = c_{ij}=:c_\alpha, \ \alpha=(ij) \in \mathcal{M},
\label{rigidity_equations}
\end{equation}
where $\mathbf{p}_i$ is the $3 \times 1$ position vector of node $i$.  Solving this set of $M$ nonlinear equations directly is usually infeasible for anything but very small systems.  An alternative approach is \emph{Infinitesimal Rigidity}, which considers infinitesimal violations of the equilibrium conditions of~\eqref{rigidity_equations}. 

Taking the derivative of both sides of~\eqref{rigidity_equations} with respect to time $t$ for all constrained pairs, we get:
\begin{equation}
(\mathbf{p}_i - \mathbf{p}_j) \cdot (\mathbf{u}_i - \mathbf{u}_j) = 0, \ (ij) \in \mathcal{M},
\end{equation}
with $\mathbf{u}_i = d \mathbf{p}_i/dt$.  We then expand out the brackets:
\begin{equation}
(\mathbf{p}_i - \mathbf{p}_j)\mathbf{u}_i - (\mathbf{p}_i - \mathbf{p}_j)\mathbf{u}_j = 0,
\end{equation}
and rewrite in vector form:
\begin{equation}
\mathbf{R}\mathbf{u} = \mathbf{0}.
\end{equation}
The $M \times N d$ matrix $\mathbf{R}$ is called the \emph{rigidity matrix} and each row represents a single constraint.  For example, a three node system with each pair of nodes joined by an edge would have the rigidity matrix~\cite{JoaoThesis}:  
\begin{equation}
\mathbf{R} =
\begin{pmatrix} \mathbf{p}_1 - \mathbf{p}_2, & \mathbf{p}_2 - \mathbf{p}_1, & \mathbf{0} \\
\mathbf{0}, & \mathbf{p}_2 - \mathbf{p}_3, & \mathbf{p}_3 - \mathbf{p}_2 \\
\mathbf{p}_1 - \mathbf{p}_3, & \mathbf{0}, & \mathbf{p}_3 - \mathbf{p}_1
\end{pmatrix}
\label{eq:rigidity_definition}
\end{equation}
The infinitesimal rigidity properties follow from examining the nullspace of
$\mathbf{R}$. Hence these properties are an intrinsic property of the structure, and are independent of the environment or the friction terms.  From~\eqref{eq:rigidity_definition}, it follows immediately that the geometric incidence matrix is a scaled version of $\mathbf{R}$.  

\section{Algorithm for rigid cluster decomposition using infinitesimal rigidity}
\label{app:Joao_algo}

The following algorithm was introduced in~\cite{JoaoThesis} and is summarised here for completeness. We use it to obtain the rigid clusters shown in Fig.~\ref{PDK1_infrig}.

\begin{figure}[!htb]
\centering
\includegraphics[width=0.42\textwidth]{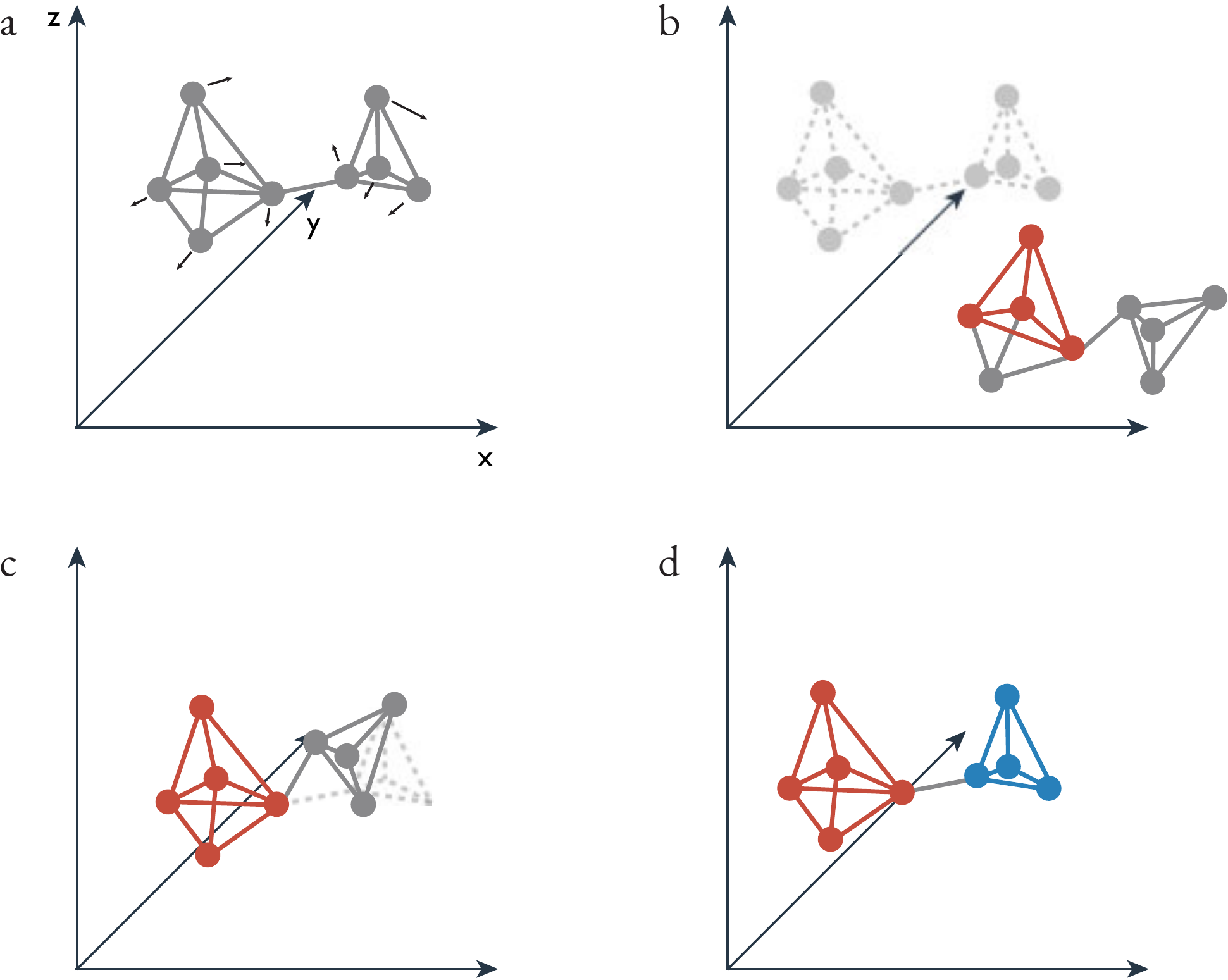}
\label{cluster_decomp}
\caption[Rigid cluster decomposition algorithm using infinitesimal rigidity]{Rigid cluster decomposition algorithm using infinitesimal rigidity~\cite{JoaoThesis}.  For each trivial infinitesimal motion, such as the one in (a), the atoms are moved by a small distance along each $3N \times 1$ vector to a new position b).  A rigid tetrahedron of atoms is selected in the new position then in c) this is moved back to its original position.  Any atoms that also return to their original position at the same time (for all infinitesimal motions) are part of the same cluster. d) The process is repeated until all atoms are clustered into rigid regions or are assigned as floppy.}
\end{figure}

The steps of the algorithm are as follows (see Fig.~\ref{cluster_decomp}):
\begin{enumerate}
  \item Identify a set of 4 atoms, $\mathcal{T}$, that form a fully connected tetrahedron.
  \item Translate the coordinate frame to the centre of the set $\mathcal{T}$.
  \begin{equation*}
  \mathbf{p}_k := \mathbf{p}_k - \frac{1}{4}\sum_{k \in \mathcal{T}} \mathbf{p}_k
  \end{equation*}
  \item Transform the three coordinate axes so that they correspond to the principle axes of the set $\mathcal{T}$:
  \begin{equation*}
  \mathbf{p}_k := \mathbf{S} \, \mathbf{p}_k \\
  \end{equation*}
  $\mathbf{S}$ is the rotation matrix whose rows are the eigenvectors of the matrix $\mathbf{I}$:
  \begin{align*}
  \mathbf{I}_{\alpha \beta} = \sum_{k \in T}(|\mathbf{p}_k|\delta_{\alpha \beta} &- \mathbf{p}_{k\alpha}\mathbf{p}_{k\beta}), 
  \\ \text{where} \ \ (\alpha, \beta) &= (x,y,z) 
  \end{align*}
  \item Generate the trivial motions in this new coordinate frame: three rotations $(\mathbf{r}_k^x, \mathbf{r}_k^y, \mathbf{r}_k^z)$ and three translations $(\mathbf{t}_k^x, \mathbf{t}_k^y, \mathbf{t}_k^z)$ for all of the atoms of the structure.
  \begin{equation*}
  \mathbf{r}_k^{\alpha} = \mathbf{p}_k \times \widehat{\mathbf{e}^{\alpha}}; \ \ \ \mathbf{t}_k^{\alpha} = \widehat{\mathbf{e}^{\alpha}}
  \end{equation*}
  \item Transform the trivial motions back into the starting coordinate frame:
  \begin{equation*}
	\mathbf{r}_k^{\alpha} := \mathbf{S}^T \mathbf{r}_k^{\alpha}; \ \ \  \mathbf{t}_k^{\alpha} := \mathbf{S}^T \mathbf{t}_k^{\alpha}
  \end{equation*}
  \item Compile the trivial motions for each atom into column vectors so we have three 3N-dimensional translations $\mathbf{t}^{\alpha}$ and three rotations $\mathbf{r}^{\alpha}$.  Normalise each of these trivial motions:
  \begin{equation*}
  \mathbf{r}^{\alpha} := \frac{\mathbf{r}^{\alpha}}{|\mathbf{r^{\alpha}_T}|}; \ \ \ \mathbf{t}^{\alpha} := \frac{\mathbf{t}^{\alpha}}{|\mathbf{t^{\alpha}_T}|}
  \end{equation*}
  using the magnitude of the 12-dimensional vectors associated with the set $\mathcal{T}$.  Now the set of six 12-dimensional trivial motions of the set $\mathcal{T}$ are orthonormal.
  \item The set of displacements of each of the atoms relative to the set $\mathcal{T}$ can then be calculated by returning the set $\mathcal{T}$ to its initial position:
  \begin{equation*}
  \Delta \mathbf{p}^{\gamma} = \mathbf{q}^{\gamma} - \sum_{\alpha}(\mathbf{q}_T^{\gamma} \cdot \mathbf{r}(t)^{\alpha})r^{\alpha} - 
  \sum_{\alpha}(\mathbf{q}_T^{\gamma} \cdot \mathbf{t}_T^{\alpha})\mathbf{t}^{\alpha}
  \end{equation*}
  where we now use $\gamma$ additionally index over the set of trivial motions. 
  \item For each atom, calculate its absolute displacement in space away from its initial position due to the infinitesimal motions.  If the maximum displacement of the atom over the entire set of infinitesimal motions is below a chosen small threshold value then we say that atom is part of the same rigid cluster as the set $\mathcal{T}$:
  \begin{equation*}
  \underset{\gamma}{\text{max}}|\Delta \mathbf{p}_k^{\gamma}| < \delta
  \end{equation*}
  where $\delta$ is a cutoff to account for floating point rounding error.  Here we use $\delta= ~10^{-4}$.
\end{enumerate}

\section{Edge displacement decays exponentially with distance}
\begin{figure}[h]
\centering
\includegraphics[width=0.45\textwidth]{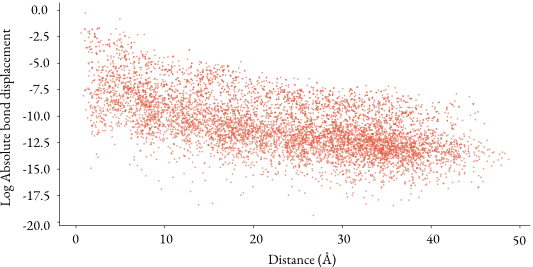}
\caption{The log absolute extension of interactions decreases linearly as a function of distance from the allosteric source site with slope -0.142 (correlation coefficient = -0.603, standard error = 0.0022), i.e., the effect of the perturbation decays exponentially away from the allosteric site.}
\label{fig:exponential_decay}
\end{figure}

In all of the protein structures studied, it was found that the absolute extension of the springs decreased exponentially with increasing distance of the spring from the perturbation site, as shown in Fig.~\ref{fig:exponential_decay}.

\end{document}